\documentclass[aps,pra,showpacs,twoside,twocolumn,10pt]{revtex4-1}
\usepackage[colorlinks=true, citecolor=red, urlcolor=blue ]{hyperref}
\usepackage{appendix}
\usepackage{epsfig,newlfont,amssymb,amsfonts,amsmath,bm,subfigure,palatino,mathtools,amsthm,braket,times,soul,enumitem,color}
\usepackage[normalem]{ulem}
\usepackage{graphicx}
\usepackage{float}
\usepackage{float}
\usepackage{amsmath}

\begin{document}
\title{Tight Product Monogamy Inequality for Entanglement}
\author{Ida Mishra, Arun K Pati and Sohail}
\affiliation{Harish-Chandra Research Institute, HBNI, Chhatnag Road, Jhunsi, Allahabad - 211 019, India}

\begin{abstract}
Quantum entanglement for multiparty system has a unique feature when it comes to sharing its property among various subsystems. This is famously
stated as the monogamy of entanglement. The traditional monogamy of concurrence for tripartite system was proved in a sum form. Recently, it was found that concurrence also respects a monogamy in the product form. Here, we prove a tight monogamy relation in the product form for the concurrence of pure tripartite systems. We illustrate our relation with several examples, including the canonical three qubit states, where this monogamy relation is saturated.
\end{abstract}
\maketitle
\section{Introduction}
Quantum entanglement \cite{Tr1, Tr2, Tr3} is one of the most weirdest feature of quantum mechanics which has fundamental as well as
technological applications in the field of Quantum Information and
Computation \cite{Tr4}. Its beauty is far beyond being just a paradoxical
notion, rather it acts as a quantum resource \cite{Tr5} in the emerging
quantum technologies. Entangled state in its simplest form, such as an
Einstein-Podolsky-Rosen (EPR) pair, can be used to perform tasks like
teleportation \cite{Tr6}, remote state preparation \cite{Tr7}, quantum
cryptography \cite{Tr8}, super-dense coding \cite{Tr9} and entanglement
swapping \cite{Tr10,Tr11} and many more. Similarly, more complex form of 
multi-qubit entangled states, enable tasks such as entanglement assisted
error correction \cite{Tr12}, one-way quantum computation \cite{Tr13} and
as tomographic resources \cite{Tr14}. Understanding and quantifying
entanglement in multipartite states are active areas of research in
Quantum Information Theory.

Considerable amount of literature has been focused for understanding the
notion of sharing of entanglement between multiparty states. It was found
that, unlike classical correlation, quantum entanglement cannot be freely
shared among many parties. This is paraphrased as the monogamy of quantum
entanglement. One extreme form of this monogamy suggests that if qubit $A$
and qubit $B$ are in a maximally entangled state (for example in one of
the Bell states), then none of the qubit of this pair can be  entangled
with third qubit $C$. There can be less restrictive form of this
entanglement sharing which was first proved by Coffman, Kundu and
Wootters through the monogamy inequality in the additive form \cite{Tr15}.
Over last several years, the monogamy inequality has been generalised for more than bipartite systems, higher
dimensional systems and other correlations measures (see for example \cite{Tr16}).
Recently, Fei{\it  et. al} \cite{Tr17} proposed a product monogamy
inequality which is different from the original monogamy inequality.  In
this paper, we prove a tighter monogamy inequality in the product form
which gives a better understanding of entanglement distribution bound for
several tripartite systems \cite{Tr18}. The product monogamy that was
obtained in Ref. \cite{Tr17} is a special case of the tighter product
monogamy inequality proved here.

The paper is organised as follows. In Section II, we recapitulate the
basic definition of the concurrence and prove the tighter product monogamy
inequality. In Section III, we illustrate the new inequality for several
tripartite systems such as the superposition of a Bell state and product
state, %Zero-Sum-Amplitude (ZSA) entangled states and .... 
We also present examples where the tighter product monogamy saturates the inequality for canonical three qubit states.
Finally, we conclude our paper in Section IV.

\section{Tighter Product Monogamy Inequality}

For the sake of completeness, let us start by defining the concurrence \cite{Tr19} for a general bipartite state $\rho_{AB}$ which may be
pure or mixed. Our formula for concurrence makes use of spin-flip operation \cite{Tr20, Tr21}, which can be
applied to states of bipartite qubit states. For a pure state of a single qubit, the spin flip density operator,
which we denote by a tilde, is defined by
\begin{center}
$|\tilde\psi\rangle=\sigma_{y}|\psi^*\rangle$,
\end{center}
where $|\psi^*\rangle$ is the complex conjugate of $|\psi\rangle$ when expressed in the computational basis. This is time reversal operation for a qubit which amount to reversing the direction of the spin. If we see the
action of spin flip operator on a qubit, we can see that it transforms the input qubit to its orthogonal state. Let the initial state of the qubit is given by
\begin{center}
$\rho=\frac{1}{2}(I+\Vec{n}.\Vec{\sigma})$.
\end{center}
The spin flipped qubit is now given by 
\begin{center}
$\rho \rightarrow \sigma_{y}\rho^*\sigma_{y}=\frac{1}{2}(I-\Vec{n}.\Vec{\sigma}),$
\end{center}
where $\Vec{n}$ is the Bloch vector and $\Vec{\sigma}$ is the Pauli matrix. 
%Similarly, the spin-flip on a two-qubit state transforms it as
%\begin{center}
%$\rho\rightarrow\rho^*\equiv\sigma_{y}^{\otimes2}\otimes\rho^*_{AB}\otimes\sigma_{y}^{\otimes2}$.
%\end{center}
%here $\sigma_{y}^n$ is n time tensor product of $\sigma_{y}$.
Similarly, we can define the spin flipped density matrix for any two qubit state. Let $\rho_{AB}$ be density matrix of a pair of qubits $A$
and $B$ and $\tilde{\rho}_{AB}$ be spin flipped density matrix \cite{Tr22}
which is defined as
\begin{center}
$\tilde{\rho}_{AB}=(\sigma_{y}\otimes\sigma_{y})\rho^*_{AB}(\sigma_{y}\otimes\sigma_{y})$.
\end{center}
Here, asterisk is complex conjugation with respect to the computational basis
{$|00\rangle,|01\rangle,|10\rangle,|11\rangle$} and $\sigma_{y}$ is the
Pauli spin matrix along the $y$-direction. %expressed as  expressed in the same basis
%\begin{centre}
 % \begin{pmatrix}
% 0& - i\\
% i  &0
%\end{pmatrix}
%\end{centre}.
The matrix $\rho_{AB}\tilde{\rho}_{AB}$, even though is
non-Hermitian, has only real and non-negative eigenvalues because both $\rho_{AB}$ and $\tilde{\rho}_{AB}$ are positive operators. 
Let $\lambda_{1},\lambda_{2},\lambda_{3}$ and $\lambda_{4}$ be square root of
eigenvalues of $\rho_{AB}\tilde{\rho}_{AB}$ in decreasing numerical order. Then the concurrence for the density matrix $\rho_{AB}$ is defined as
\begin{center}
  $ C_{AB}=max(\lambda_{1}-\lambda_{2}-\lambda_{3}-\lambda_{4},0).$
\end{center}
The concept is simpler for describing 2-qubit pure state entanglement for which concurrence is defined as
\begin{center}
 $C(\Psi_{AB})=|\langle\Psi_{AB}|\tilde{\Psi}^*_{AB}\rangle|=2\sqrt{det\rho_{A}}$
\end{center}
where $\rho_{A}$ is reduced density matrix of system $A$. It has a physical meaning for any 2-qubit pure state as it suggests how close the spin-flipped state is to the original state. The concurrence ranges from $0$ to $1$, i.e., for $C=0$ and $C=1$ correspond to
unentangled and maximally entangled state, respectively.
In the present paper, we are concerned about pair of qubits, hence we define concurrence for such state, not for higher dimensional systems \cite{Tr23,Tr24}.
\\
\\
Monogamy is obeyed by several non-classical correlations like quantum discord \cite{Tr25,Tr26}, Bell non-locality \cite{Tr27,Tr28}, EPR steering \cite{Tr29,Tr30} etc. but for this paper we are interested in the monogamy of entanglement  \cite{Tr31,Tr32,Tr33,Tr34,Tr36,Tr37}. The traditional form of the monogamy of entanglement is captured by the following inequality
\begin{eqnarray}
  C^2_{AB}+ C^2_{AC} \leq  C^2_{A(BC)}.
\end{eqnarray}
This suggest that sum of entanglement shared between $AB$ and $AC$ cannot be arbitrary. It is restricted by the entanglement shared across $A$ and $BC$ as a whole. The product monogamy of concurrence is proposed by Fei et. al. for three-qubit pure states in Ref.\cite{Tr25} which is given by
\begin{equation}
C^2_{A(BC)}\geq 2\sqrt{C^2_{AB}C^2_{AC}+\frac{\tau^2_{ABC}}{4}}, \label{fei_inequality}
\end{equation}
where $C_{A(BC)}$ is the concurrence between qubit $A$ and the pair $BC$, $C_{AB}$ and $C_{AC}$ are the concurrences for qubits $AB$ and  $AC$, respectively. Here, $\tau_{ABC}$ is the entanglement distributed between $ABC$ in a pure state and also known as the residual entanglement. Explicitly, it is given by $\tau_{ABC} = C^2_{A(BC)} - (C^2_{AB}+ C^2_{AC} )$ and it is permutation invariant, i.e., symmetric with respect to interchanging the labels $A$, $B$ and $C$.

In what follows, we derive the product monogamy inequality which is tighter than the inequality derived by Fei.et. al \cite{Tr25}.
For any arbitrary pure state $\ket{\Psi}_{ABC}$ of 3-qubits, the concurrence $C_{AB}$ for the party $AB$ satisfies the following relation (see \cite{Tr24})
\begin{equation}
C^2_{AB}=Tr(\rho_{AB}\tilde{\rho}_{AB})-2\lambda_1\lambda_2,
\label{concurence for pure state}
\end{equation}
where $\rho_{AB}$ is the reduced density matrix of pair $AB$ which is obtained by tracing out $C$ from the composite pure state and $\lambda_1, \lambda_2$ being the square root of two non zero eigenvalues of $\rho_{AB}\tilde\rho_{AB}$. Since, $\lambda_1\lambda_2=\frac{\tau_{ABC}}{4}$, Eq.~(\ref{concurence for pure state}) becomes
\begin{equation}
C^2_{AB}=Tr(\rho_{AB}\tilde{\rho}_{AB})-\frac{\tau_{ABC}}{2}. \label{aaa}
\end{equation}

Similarly for $AC$, we have 
\begin{equation}
C^2_{AC}=Tr(\rho_{AC}\tilde{\rho}_{AC})-\frac{\tau_{ABC}}{2}. \label{bbb}
\end{equation}
Now, using Eq.~(\ref{aaa}) and ~(\ref{bbb}) we can compute $C^2_{AB}C^2_{AC}$ as follows:
\begin{eqnarray*} 
C^2_{AB}C^2_{AC} & = (Tr(\rho_{AB}\tilde{\rho}_{AB})-\frac{\tau_{ABC}}{2})(Tr(\rho_{AC}\tilde{\rho}_{AC})-\frac{\tau_{ABC}}{2}) \nonumber \\   
& = Tr(\rho_{AB}\tilde{\rho}_{AB})Tr(\rho_{AC}\tilde{\rho}_{AC})-\frac{\tau_{ABC}}{2}( Tr(\rho_{AB}\tilde{\rho}_{AB})\\&+Tr(\rho_{AC}\tilde{\rho}_{AC})) + \frac{\tau^2_{ABC}}{4}\nonumber \\
& = Tr(\rho_{AB}\tilde{\rho}_{AB})Tr(\rho_{AC}\tilde{\rho}_{AC})\\&+\frac{1}{4}[\tau^2_{ABC} - 2\tau_{ABC}(Tr(\rho_{AB}\tilde{\rho}_{AB})+Tr(\rho_{AC}\tilde{\rho}_{AC}))]
\nonumber \\
& = Tr(\rho_{AB}\tilde{\rho}_{AB})Tr(\rho_{AC}\tilde{\rho}_{AC})\\&+\frac{1}{4}[(Tr(\rho_{AB}\tilde{\rho}_{AB})+Tr(\rho_{AC}\tilde{\rho}_{AC})-\tau_{ABC})^2\\&
-(Tr(\rho_{AB}\tilde{\rho}_{AB})+Tr(\rho_{AC}\tilde{\rho}_{AC}))^2]
\nonumber \\
& = Tr(\rho_{AB}\tilde{\rho}_{AB})Tr(\rho_{AC}\tilde{\rho}_{AC})\\&+\frac{1}{4}[(C^2_{AB}+C^2_{AC})^2 - C^4_{A(BC)}]
\nonumber \\
& \geq Tr(\rho_{AB}\tilde{\rho}_{AB})Tr(\rho_{AC}\tilde{\rho}_{AC})\\&+\frac{1}{4}[4C^2_{AB}C^2_{AC} - C^4_{A(BC)}], \label{eq:five}
\end{eqnarray*}
where the equality in $5^{th}$ step is computed by the substitution from Eq.~(\ref{aaa}) and (\ref{bbb}) and using $ Tr(\rho_{AB}\tilde{\rho}_{AB})+Tr(\rho_{AC}\tilde{\rho}_{AC})=C^2_{A(BC)}$ while the inequality is based on the arithmetic geometry mean inequality $a^2+b^2\geq 2ab$ for all $a,b \geq 0$.
\\After simplification of the above inequality we have
\begin{center}
 $C^4_{A(BC)}\geq 4Tr(\rho_{AB}\tilde{\rho}_{AB})Tr(\rho_{AC}\tilde{\rho}_{AC})$.
\end{center}
Substituting $Tr(\rho_{AB}\tilde{\rho}_{AB})$ and $Tr(\rho_{AC}\tilde{\rho}_{AC})$ from equation eq.~(\ref{aaa}) and (\ref{bbb}), we have the following inequality
\\
%\begin{center}
%  $C^4_{A(BC)}\geq 4(C^2_{AB}+\frac{\tau_{ABC}}{2})(C^2_{AC}+\frac{\tau_{ABC}}{2})$
%\end{center}
\begin{equation}
  C^2_{A(BC)}\geq 2\sqrt{(C^2_{AB}+\frac{\tau_{ABC}}{2})(C^2_{AC}+\frac{\tau_{ABC}}{2})}.
  \label{product_monogamy}
\end{equation}
\\ 
The above product monogamy inequality~(\ref{product_monogamy}) is stronger than the inequality~(\ref{fei_inequality}) proposed by Fei {\it etal} \cite{Tr25}. This is the central result of the paper.
The tight monogamy inequality for the concurrence can shed some new light on the nature of entanglement distribution between tripartite pure states. In the sequel, we illustrate the product monogamy inequality 
with several examples.

\section{STATES SATURATING TIGHTER PRODUCT MONOGAMY INEQUALITY}
\label{sec:sec3}
States saturating any given inequality are of great interest for several applications. %called minimum uncertainty states by general nomenclature; by lowest uncertainty we mean that these states satisfy eq. $(5)$ as equality. These states are of great interest for several application purposes and they not always easy to find.
Here, we show some states that saturates the tighter monogamy inequality~(\ref{product_monogamy}).

First, consider the famous Greenberger-Horne-Zeilinger (GHZ) state $|\Psi\rangle_{ABC} = \frac{1}{\sqrt 2} (| 000 \rangle_{ABC} +  |111 \rangle_{ABC}) $.
The GHZ state has the important features that qubits in each pair are not entangled, rather they are classically correlated. However, each qubit is entangled with any other two qubits.
For this state the concurrence of qubit $A$ with the subsystems $BC$ is given by  $C^2_{A(BC)} =1$.
Also, it is easy to check that the pairwise concurrences are zero, i.e., $C^2_{AB} =0$ and $C^2_{AC} =0$ and hence the residual entanglement $\tau_{ABC} = 1$. Therefore, the tight product monogamy inequality is actually saturated for the GHZ state.

Let us consider the following tripartite pure entangled state
\begin{equation}
  |\Psi\rangle_{ABC} = \sqrt{p_1}|\Psi^-\rangle |0\rangle +\sqrt{p_2}|00\rangle|1\rangle, \nonumber\\ \label{saturating state 1}
\end{equation}
 where $|\Psi^-\rangle= \frac{1}{\sqrt{2}}(\ket{01}-\ket{10})$ is a Bell state and $p_1$, $p_2$ are positive real numbers such that $p_1 + p_2=1$.
The reduced state of the pair AB is given by 
\begin{align}
\rho_{AB}= p_1|\Psi^-\rangle\langle \Psi^-|+p_2|00\rangle\langle 00|.
\label{state 1}
\end{align}
%\iffalse
%Matrix form of density matrix of pair AB is
%\begin{equation}
% \rho_{AB}= 
% \begin{pmatrix}
% p_2&0&0&0\\
% 0&p_1/2&-p_1/2&0\\
% 0&-p_1/2&p_1/2&0\\
% 0&0&0&0
% \end{pmatrix}
% \end{equation}
%\fi
The reduced state is a special case of a mixed maximally entangled state \cite{Tr38}. These class of states have special feature whose entanglement cannot be enhanced by application of any local and non-local
unitary transformation. Now, the spin flipped density matrix of pair $AB$ is given by
\begin{align}
\tilde{\rho}_{AB} & = (\sigma_{y}\otimes\sigma_{y})\rho^*_{AB}(\sigma_{y}\otimes\sigma_{y}) \nonumber\\
& = 
\begin{pmatrix}
0&0&0&0\\[0.5em]
0&\frac{p_1}{2}&-\frac{p_1}{2}&0\\[0.5em]
0&-\frac{p_1}{2}&\frac{p_1}{2}&0\\[0.5em]
0&0&0&p_2
\end{pmatrix}. \label{spin fliped 1}
\end{align}
\\Using eq.~(\ref{state 1}) and eq.~(\ref{spin fliped 1}) we get
\begin{center}
$\rho_{AB}\tilde{\rho}_{AB}=
\begin{pmatrix}
0&0&0&0\\[0.5em]
0&\frac{p_1^2}{2}&-\frac{p_1^2}{2}&0\\[0.5em]
0&-\frac{p_1^2}{2}&\frac{p_1^2}{2}&0\\[0.5em]
0&0&0&0
\end{pmatrix}$.
\end{center}
The square root of eigenvalues of $\rho_{AB}\tilde{\rho}_{AB}$ are $\lambda_1=p_1$ and $\lambda_2= \lambda_3=\lambda_4=0$. This gives
the concurrence of $\rho_{AB}$ as
\begin{equation}
\begin{split}
C(\rho_{AB})& = max(\lambda_1-\lambda_2-\lambda_3-\lambda_4,0) = p_1 \label{Concurrence_AB}
\end{split}.
\end{equation}
Similarly, the reduced density matrix and spin flipped density matrix for the pair $AC$ are, respectively, given by
\begin{align}
\rho_{AC} & = Tr_{B}(\rho_{ABC})  \nonumber \\
& = \frac{p_1}{2}(|00\rangle\langle00|+|10\rangle\langle10|)+p_2(|01\rangle\langle01|) \nonumber \\
& \quad-\sqrt\frac{p_1p_2}{2}(|10\rangle\langle01|+|01\rangle\langle10|) \nonumber \\
& = 
\begin{pmatrix}
\frac{p_1}{2}&0&0&0\\[0.5em]
0&p_2&-\frac{\sqrt{p_1p_2}}{\sqrt{2}}&0\\[0.5em]
0&-\frac{\sqrt{p_1p_2}}{\sqrt{2}}&\frac{p_1}{2}&0\\[0.5em]
0&0&0&0
\end{pmatrix} \label{state of AC}
\end{align}
and
\begin{align}
\tilde{\rho}_{AC} & = (\sigma_{y}\otimes\sigma_{y})\rho^*_{AC}(\sigma_{y}\otimes\sigma_{y})  \nonumber \\
&=
\begin{pmatrix}
0&0&0&0\\[1em]
0&\frac{p_1}{2}&-\frac{\sqrt{p_1p_2}}{\sqrt{2}}&0\\[1em]
0&-\frac{\sqrt{p_1p_2}}{\sqrt{2}}&p_2&0\\[1em]
0&0&0&\frac{p_1}{2}
\end{pmatrix}. \label{spin fliped AC}
\end{align}
Using Eq.~(\ref{state of AC}) and Eq.~(\ref{spin fliped AC}) we obtain
\begin{center}
$\rho_{AC}\tilde{\rho}_{AC}=
\begin{pmatrix}
0&0&0&0\\[1em]
0&p_1p_2&-\sqrt{2p_1p_2^3}&0\\[1em]
0&-\frac{\sqrt{p_1^3p_2}}{\sqrt{2}}&p_1p_2&0\\[1em]
0&0&0&0
\end{pmatrix}$.
\end{center}
Now, the square root of eigenvalues of $\rho_{AC}\tilde{\rho}_{AC}$ are given by $\sqrt{2p_1p_2}, 0, 0, 0$. Thus, we have the concurrence of $\rho_{AC}$ as
\begin{equation}
\begin{split}
C(\rho_{AC})& = \sqrt{2p_1p_2}. \label{Concurrence_AC}
\end{split}
\end{equation}
Since $\rho_{ABC}$ is pure state, the concurrence of composite state in $A:BC$ bipartition can easily be evaluated. It is given by
\begin{equation}
   C_{A(BC)}=\sqrt{2(1-Tr(\rho^2_{A}))}. \label{pure_concurrence}
\end{equation}
%Now, the reduced density matrix for sub-system $A$ is,%
%\iffalse
%\begin{align}
%\rho_{A} & = Tr_{BC}\left(\rho_{ABC}\right) %\nonumber\\
%\rho_{A} & = %(\frac{p_1}{2}+p_2)|0\rangle\langle %0|+\frac{p_1}{2}|1\rangle\langle 1| \nonumber\\
%& = 
%\begin{pmatrix}
%p_1/2+p_2&0\\
%0&p_1/2
%\end{pmatrix}\nonumber
%\end{align}
%\fi
%So,
%\iffalse
%\begin{center}
%$\rho^2_{A}=
%\begin{pmatrix}
%(p_1/2+p_2)^2&0\\
%0&(p_1/2)^2
%\end{pmatrix}$
%\end{center}
%We get the trace of $\rho^2_{A}$ as,
%\begin{equation*}
 %   Tr(\rho^2_{A})=(\frac{p_1}{2}+p_2)^2+(\frac%{p_1}{2})^2.
%\end{equation*}
%\fi
%\\We now calculate $C_{A(BC)}$ using Eq.~(\ref{pure_concurrence}) as the following,%
\begin{align}
C_{A(BC)} = \sqrt{p_1(p_2+1)}.   \label{Concurrence_(ABC)}
\end{align}

%Using eq.~ (\ref{Concurrence_AB}),~ (\ref{Concurrence_AC}) and (\ref{Concurrence_(ABC)}, we can check that $\rho_{(ABC)}$ saturates the CKW monogamy inequality
%\begin{equation}
 %   C^2_{AB}+C^2_{AC}\leq C^2_{A(BC)} \label{CKW inequality}
%\end{equation}
%for all values of $p_1, p_2$.% Hence $\rho_{ABC}$ saturates CKW monogamy inequality.%

Now, we check whether this pure tripartite state saturates the tighter product  monogamy inequality~(\ref{product_monogamy}) or not. Remember that $ \tau_{ABC}=4\lambda_1\lambda_2$ where
$\lambda_1,\lambda_2$ are square root of eigenvalues of $\rho_{(AB/AC)}\tilde{\rho}_{(AB/AC)}$. As $\lambda_2=0$ for both pair $AB$ and pair $AC$, so  $\tau_{ABC}=0$. Hence, tighter product monogamy inequality reduces to 
\begin{equation}
 C^2_{A(BC)}\geq 2\sqrt{(C^2_{AB})(C^2_{AC})} \label{reduced inequality}
\end{equation}
Using ~(\ref{Concurrence_AB}),~(\ref{Concurrence_AC}) and ~(\ref{Concurrence_(ABC)}) in the inequality ~(\ref{reduced inequality}), it is not difficult to observe that it is saturated by the state ~(\ref{saturating state 1}) when we have $\sqrt{p_1}=\sqrt{2p_2}$.

%Similar results hold for all the states listed below. 
%\iffalse
%\begin{center}
%1) $|\psi\rangle_{ABC}=\sqrt{p_1}|\psi^-\rangle %|0\rangle +\sqrt{p_2}|11\rangle|1\rangle$
%    
%    \vskip 0.2cm
%2) $|\psi\rangle_{ABC}=\sqrt{p_1}|\phi^-\rangle %|0\rangle +\sqrt{p_2}|01\rangle|1\rangle$
% \vskip 0.2cm
%3) $|\psi\rangle_{ABC}=\sqrt{p_1}|\phi^-\rangle %|0\rangle +\sqrt{p_2}|10\rangle|1\rangle$
% \vskip 0.2cm
%4) $|\psi\rangle_{ABC}=\sqrt{p_1}|\psi^+\rangle %|0\rangle +\sqrt{p_2}|00\rangle|1\rangle$
% \vskip 0.2cm
%5) $|\psi\rangle_{ABC}=\sqrt{p_1}|\psi^+\rangle %|0\rangle +\sqrt{p_2}|11\rangle|1\rangle$
% \vskip 0.2cm
%6) $|\psi\rangle_{ABC}=\sqrt{p_1}|\phi^+\rangle %|0\rangle +\sqrt{p_2}|01\rangle|1\rangle$
% \vskip 0.2cm
%7) $|\psi\rangle_{ABC}=\sqrt{p_1}|\phi^+\rangle %|0\rangle +\sqrt{p_2}|10\rangle|1\rangle$
%\end{center}
%where $p_1,p_2$ are positive real numbers with %$p_1+p_2=1$.
%\\
%Now consider the following states which are %slightly different from the states discussed %above,
%\begin{center}
%    9) $|\psi\rangle_{ABC}=\sqrt{p_1}|011\rangl%e +\sqrt{p_2}|101\rangle+\sqrt{p_3}|110\rangle$
%    \\
%10) $|\psi\rangle_{ABC}=\sqrt{p_1}|100\rangle %+\sqrt{p_2}|010\rangle+\sqrt{p_3}|001\rangle$
%\end{center}
%\fi
%where $p_1,p_2,p_3$ are positive real numbers %such that $p_1+p_2+p_3=1$.
%\\ It can easily be shown that the above two %states saturate CKW monogamy %inequality~(\ref{CKW inequality}) for all %values of $p_1, p_2, p_3$ and saturate tighter %product monogamy %inequality~(\ref{product_monogamy}) when %$p_2=p_3$ and all values of $p_1$ compatible %with the normalization condition.

Next, we explore how tight our inequality (\ref{product_monogamy}) is compared to the one derived by Fei. et al. (\ref{fei_inequality}) using the canonical form of three qubit pure state (see for instance \cite{Tr39,Tr40}). Any generic state shared by $3$-parties belonging to $C^2\otimes C^2\otimes C^2$ can be written as
\begin{align}
 \Psi =p_1e^{i\theta}|000\rangle+p_2|001\rangle +p_3|100\rangle +p_4|110\rangle +p_5 |111\rangle, \label{cc}
\end{align}
where $p_i$ are real and positive and $0 \leq \theta < \pi$. These class of states play important role as we do not have the Schmidt decomposition theorem for tripartite systems. 

Now, the reduced density matrix for the pair $AB$ is given by
\begin{align}
\rho_{AB} & = Tr_{C}(\rho_{ABC})  \nonumber \\
& = 
\begin{pmatrix}
p_1^2+p_2^2&0&p_1p_3e^{i\theta}&p_2p_5+p_1p_4e^{i\theta}\\[0.5em]
0&0&0&0\\[0.5em]
p_1p_3e^{-i\theta}&0&p_3^2&p_3p_4\\[0.5em]
p_2p_5+p_1p_4e^{-i\theta}&0&p_3p_4&p_4^2+p_5^2
\end{pmatrix}. \label{state of AB-can}
\end{align}
The spin flipped density matrix for the pair $AB$ is given by
\begin{align}
\tilde{\rho}_{AB} & =(\sigma_{y}\otimes\sigma_{y})\rho_{AB}(\sigma_{y}\otimes\sigma_{y})  \nonumber \\
& = 
\begin{pmatrix}
p_4^2+p_5^2&-p_3p_4&0&p_2p_5+p_1p_4e^{-i\theta}\\[0.5em]
-p_3p_4&p_3^2&0&-p_1p_3e^{-i \theta}\\[0.5em]
0&0&0&0\\[0.5em]
p_2p_5+p_1p_4e^{i\theta}&-p_1p_3e^{i \theta}&0&p_1^2+p_2^2
\end{pmatrix}. \label{ABtilde-can}
\end{align}
Using eq.(\ref{state of AB-can}) and (\ref{ABtilde-can}), we can get density matrix of $\rho_{AB}\tilde{\rho}_{AB}$ and $C^2_{AB}$ as
\begin{align}
\nonumber
    C^2_{AB}= 4(p_2^2p_5^2 + p_1^2p_4^2cos^2\theta + 2p_1p_2p_4p_5cos\theta)
\end{align}
Similarly, the reduced density matrix for the pair $AC$ is given by
\begin{align}
\rho_{AC} & = Tr_{A}(\rho_{ABC})  \nonumber \\
& = 
\begin{pmatrix}
p_1^2&p_1p_2e^{i\theta}&p_1p_3e^{i\theta}&0\\[0.5em]
p_1p_2e^{-i\theta}&p_2^2&p_2p_3&0\\[0.5em]
p_1p_3e^{-i\theta}&p_3p_4&p_3^2+p_4^2&p_3p_4\\[0.5em]
0&0&p_3p_4&p_5^2
\end{pmatrix}. \label{state of AC-can}
\end{align}
The spin flipped density matrix for the pair $AC$ is given by
\begin{align}
\tilde{\rho}_{AC} & =(\sigma_{y}\otimes\sigma_{y})\rho_{AC}(\sigma_{y}\otimes\sigma_{y})  \nonumber \\
& = 
\begin{pmatrix}
p_5^2&-p_4p_5&0&0\\[0.5em]
-p_4p_5&p_3^2+p_4^2&p_2p_3&-p_1p_3e^{-i \theta}\\[0.5em]
0&p_2p_3&p_2^2&-p_1p_2e^{-i \theta}\\[0.5em]
0&-p_1p_3e^{i \theta}&-p_1p_2e^{i \theta}&p_1^2
\end{pmatrix}. \label{ACtilde-can}
\end{align}
Using eq.(\ref{state of AC-can}) and (\ref{ACtilde-can}), we can get density matrix of $\rho_{AC}\tilde{\rho_{AC}}$ and $C^2_{AC}$ as given by
\begin{align}
\nonumber
C^2_{AC}= 4p_2^2p_3^2 
\end{align}
Now, the residual entanglement $\tau_{ABC}$ is given by
%which can be mathematically calculated by multiplication of eigenstates of $\rho_{AB}\tilde{\rho_{AB}}$ or $\rho_{AC}\tilde{\rho_{AC}}$ is
\begin{eqnarray}
\nonumber
   \tau_{ABC} = 4(p_2^2p_4^2 + p_1^2 p_5^2 - 2p_1p_2p_4p_5 cos\theta)^2 
\end{eqnarray}
To find the concurrence $C_{A(BC)}$, we find the reduced density matrix of $A $ which is given by 
\begin{align}
\rho_A & = Tr_{BC}(\rho_{ABC})  
\nonumber \\
& = 
\begin{pmatrix}
p_1^2+p_2^2&p_1p_3e^{i\theta}\\[0.5em]
p_1p_3e^{-i\theta}&p_3^2+p_4^2+p_5^2 
\end{pmatrix}. \label{A-can}
\end{align}
From eq.~(\ref{A-can}), $C^2_{A(BC)}$ can be calculated and is given by 
\begin{align}
     C^2_{A(BC)}&= 2(1-Tr(\rho_{A}^2)) \nonumber\\
     &= 4 (p_1^4 - p_2^2 + p_2^4 + p_1^2 (1 - 2 p_2^2 - p_3^2)).  \nonumber
\end{align}
Now, we explore the tightness of our inequality and compare this with the Fei {\it etal} inequality via plotting it with right hand side of eq.~(\ref{product_monogamy}) and eq.~(\ref{fei_inequality}) in the same plot.
\begin{figure}[htbp]
\centerline{\includegraphics[width=3in, height=2.5in]{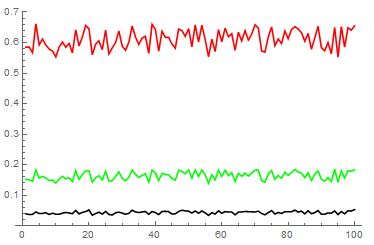}}
\caption{Red plot is for $C^2_{A(BC)}$, green and black plot are right hand side of eq.~(\ref{product_monogamy}) and eq.~(\ref{fei_inequality}) respectively where $x$-axis represents $100$ random states of the form ~(\ref{cc}). Note that the ordering of the states along $x$-axis is not important, in particular, the shapes of the curves are not important. What is important here is that the green curve is more closer to the red curve than the black curve implying that our derived inequality is more tighter than that derived by Fei. et al.}
\label{fig1}
\end{figure}
\begin{figure}[htbp]
\centerline{\includegraphics[width=3in, height=2.5in]{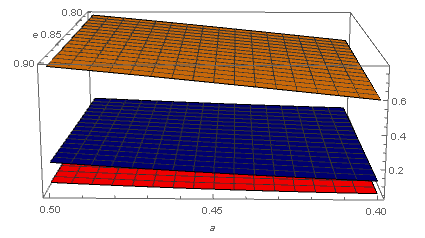}}
\caption{Top plot is for $C^2_{A(BC)}$, middle and bottom plot are right hand side of eq.~(\ref{product_monogamy}) and eq.~(\ref{fei_inequality}), respectively, where $x$-axis and $y$-axis represents $0.4 < p_1 < 0.5$ and $ 0.8 <p_5 < 0.9$ respectively with other coefficients fixed as $p_2=0.17, p_3= 0.16, p_4=0.15, \theta=0.$}
\label{fig2}
\end{figure}
Fig.~(\ref{fig1}) depicts that inequality given by Eq.~(\ref{product_monogamy}) is tighter than Eq.~(\ref{fei_inequality}). It is interesting to see $3$-d plot of the same for $100$ random states, which signifies that they do not coincide for any state. 

\begin{figure}[htbp]
\centerline{\includegraphics[width=2.5in, height=1.7in]{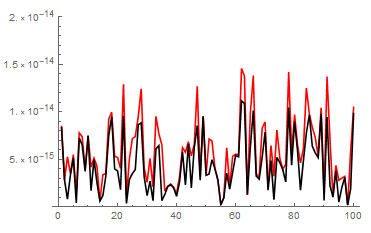} }
\caption{The red and black plots are left and right side of~(\ref{product_monogamy}), respectively. The figure is plotted for 100 random states (x-axis) of the form eq.~(\ref{cc}) with different $p_1, p_2, p_3, p_4$ and $p_5$ values.}
\label{fig3}
\end{figure}

We find that the canonical states prove to be interesting as they approach saturation for inequality~(\ref{product_monogamy}).
The canonical state above is not saturating the tighter product monogamy inequality absolutely but the the difference in LHS and RHS is almost nearing to zero as can be seen in fig.~(\ref{fig3}). All the above results can be proved for other states in the canonical class of states in the same manner as done above as they show similar behaviour.
To vivify our results, we depict near zero saturation via another canonical state as given by 
\begin{align}
\ket{ \Psi } = p_1e^{i\theta}|000\rangle + p_2|001\rangle +p_3|010\rangle +p_4|100\rangle +p_5 |111\rangle, \label{can}
\end{align}
where $p_i$ are real and positive and $0 \leq \theta < \pi$. In the sequel, we have taken $\theta=0$. We find that the concurrences across various partitions are given by
$C^2_{AB}= 4(p_3p_4-p_2p_5)^2,     C^2_{AC}= 4(p_2p_4-p_3p_5)^2$ and $ C^2_{A(BC)} = -4(p_4^2-p_5^2+p_5^4+p_4^2(-1+p_1^2+2p_5^2))$.
The residual entanglement is given by $\tau_{ABC} = 4p_5^2(4p_2p_3p_4+p_1^2p_5)^2$.

For both of these canonical three qubits states, we find that the LHS and RHS of the product monogamy inequality is almost tight. We can see from Figs $3$ and $4$ that the difference between the LHS and the RHS of the new inequality is almost close to zero, i.e., the difference is $10^{-10} $ to $10^{-14} $. Thus, the generic three qubit entangled states indeed saturate the tight product monogamy inequality.

\begin{figure}[htbp]
\centerline{\includegraphics[width=2.5in, height=1.7in]{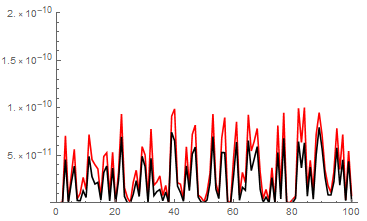}}
\caption{The red and black plots are left and right side of~(\ref{product_monogamy}), respectively. The figure is plotted for 100 random states (x-axis) of the form eq.~(\ref{can}) with different $p_1, p_2, p_3, p_4$ and $p_5$ values.}
\label{fig4}
\end{figure}

\section{CONCLUSIONS}
The monogamy of entanglement dictates how the quantum correlation is shared between multipartite systems. The monogamy nature of the concurrence was initially proved in the sum form. 
Recently,  a product monogamy was proved which is different from the original monogamy inequality.  In
this paper, we prove a tighter monogamy inequality in the product form
which gives a better understanding of entanglement distribution bound for
several tripartite systems.  The product monogamy inequality that was
obtained in Ref. \cite{Tr17} is a special case of the tighter product
monogamy inequality proved here. We have illustrated the new inequality with several examples and shown that it is indeed tight.
We have also shown that for canonical three qubit states, the product monogamy inequality is almost saturated. We hope that the tight product inequality will throw
new light on the nature of entanglement distribution in multiparticle systems.

\section{Acknowledgement}
Ida Mishra acknowledges support from HRI, Allahabad during her visit from Dec 2019 to March, 2020 for her Master Thesis project.

\end{document}